\documentclass[twocolumn,pre,showpacs]{revtex4-1}
\usepackage{graphicx}
\usepackage{amsmath}
\usepackage{amsfonts}
\usepackage{mathtools}
\usepackage{subfigure}
\usepackage[utf8]{inputenc}
\usepackage{psfrag}

\begin{document}
\title{Coherence of Biochemical Oscillations is Bounded by\\
Driving Force and Network Topology}

\author{Andre C. Barato$^{1}$ and Udo Seifert$^{2}$}
\affiliation{ $^1$ Max Planck Institute for the Physics of Complex Systems, N\"othnizer Straße 38, 01187 Dresden,Germany\\
              $^2$ II. Institut f\"ur Theoretische Physik, Universit\"at Stuttgart, 70550 Stuttgart, Germany}

\parskip 1mm
\def\A{\mathcal{A}} 
\def\R{\mathcal{R}}

\begin{abstract}
Biochemical oscillations are prevalent in living organisms. Systems with a small number of constituents
cannot sustain coherent oscillations for an indefinite time because of fluctuations in the period 
of oscillation. We show that the number of coherent oscillations that quantifies the precision of the oscillator is 
universally bounded by the thermodynamic force that drives the system out of equilibrium and by the topology of the 
underlying biochemical network of states. Our results are 
valid for arbitrary Markov processes, which are commonly used to model biochemical reactions. We apply our results 
to a model for a single KaiC protein and to an activator-inhibitor model that consists of several molecules. From a mathematical 
perspective, based on strong numerical evidence, we conjecture a universal constraint relating the imaginary and
real parts of the first non-trivial eigenvalue of a stochastic matrix.
\end{abstract}

% Explanation of PACS numbers:
% 87.16.-b: Subcellular structure and processes
% 05.70.Ln: Nonequilibrium and irreversible thermodynamics
% 05.40.-a: Fluctuation phenomena, random processes, noise, and Brownian motion

\maketitle

%========================================================================
%Introduction
%========================================================================

\section{Introduction}

Circadian rhythms \cite{gold97}, the cell cycle \cite{ferr11} and gene expression 
in somitogenesis \cite{lewi03} constitute examples of biochemical oscillations that are of central 
importance for the functioning of living systems. While older observations of biochemical 
oscillations were made with glycosis \cite{pye66},
more recent advances include the observation of 24-h oscillations of the phosphorylation level of 
the Kai proteins that  form the circadian clock of a cyanobacterium \cite{naka05,dong08}. 
Synthetically engineered genetic circuits can also display  
oscillatory behavior \cite{potv16}. On the theoretical side, 
the basic conditions for biochemical oscillations to set in are well understood  for 
deterministic rate equations that ignore fluctuations \cite{nova08}. Such rate
equations correspond to an effective description of an underlying set of chemical reactions that
is fully described by a stochastic chemical master equation.

In principle, biochemical oscillations can happen in a small system with large fluctuations in the chemical 
species that oscillates, leading to variability in the period of oscillations. Hence, 
stochastic biochemical oscillations cannot be coherent for an indefinite time. The number 
of coherent oscillations, which quantifies the precision
of the biochemical oscillator, is given by the time for which oscillations remain coherent 
divided by the period of oscillation \cite{more07,cao15}. In such a context, a relevant question is as follows: 
Given a biochemical system with significant fluctuations, what is the number of coherent oscillations that 
can be sustained?

In a recent work related to this question, Cao et al. \cite{cao15} have investigated several stochastic 
models that display biochemical oscillations. They demonstrated that this number of coherent of oscillations
increases with a larger rate of entropy production, which quantifies the free energy consumption 
of the biochemical system. Their work can be seen as part of the recent effort to understand the 
relation between a certain kind of precision and free energy consumption in biological systems, 
which include studies on kinetic proofreading \cite{qian07}, adaptation \cite{lan12},
cellular sensing \cite{meht12,bara13b,palo13,lang14,gove14,gove14a,hart15},  
information processing \cite{sart14,bara14a,bo15,ito15,mcgr17}, and cost of precision in Brownian clocks \cite{bara16}. 
In particular, we have recently shown a general relation that establishes the minimal energetic cost for 
a certain precision associated with a random variable like the output of a chemical reaction. 
This thermodynamic uncertainty relation \cite{bara15a,piet16,ging16} can be used to  infer an unknown 
enzymatic scheme in single molecule experiments \cite{bara15} and yields a bound on the efficiency of 
a molecular motor \cite{piet16b}.

In this paper, we obtain a universal bound on the number of coherent oscillations  
that can be sustained in any biochemical system that can be modelled as a Markov process 
with discrete states. This universal bound depends on the thermodynamic forces that drive the 
system out of equilibrium and on the topology of the network of states.
Our results are derived from a conjecture about the first non-trivial eigenvalue of a stochastic matrix that we 
support with thorough numerical evidence. Specifically, we obtain a bound on the ratio of the imaginary and
real parts of this eigenvalue that quantifies the number of coherent oscillations. We illustrate
our results with a model for a single KaiC protein \cite{vanz07} and with an activator-inhibitor model with several 
molecules \cite{cao15}.

The paper is organized as follows. We consider the simple case of a unicyclic network in Sec. \ref{sec2}. Our general 
bound for arbitrary multicyclic networks is formulated in Sec. \ref{sec3}. In Sec. \ref{sec4} we apply our 
results to the two models. We conclude in Sec. \ref{sec5}. In Appendix \ref{app1} we provide evidence for the 
bound for the case of unicyclic networks. The relation between the number of coherent oscillations and the Fano factor 
is discussed   in Appendix \ref{app2}. Numerical evidence for our conjecture is presented in Appendix \ref{app3}. 
Appendix \ref{app4} is dedicated to the model for a single KaiC. The relation between number of coherent 
oscillation and the entropy production in analyzed in Appendix \ref{app5}. Finally, Appendix \ref{app6} is 
dedicated to the activator-inhibitor model.

%==========================================================================
\section{Unicycic network}
%==========================================================================
\label{sec2}

As a simple model for a biochemical oscillation we start with a single enzyme $E$ with the unicyclic reaction scheme 
\begin{equation}
E_1\xrightleftharpoons[k_2^-]{k_1^+} E_2 \xrightleftharpoons[k_3^-]{k_2^+} E_3\ldots E_{N-1}\xrightleftharpoons[k_{N}^-]{k_{N-1}^+}E_N \xrightleftharpoons[k_{1}^-]{k_{N}^+}E_1,
\label{fullreaction} 
\end{equation}
where $k_i^{\pm}$ are transition rates. A generic transition from state $E_i$ to $E_{i+1}$ can represent, for example, a conformational change, binding of substrate to the enzyme
or the release of a product from the enzyme. The thermodynamic force driving this system out of equilibrium is given by the affinity \cite{seif12} 
\begin{equation}
\A\equiv \ln\prod_{i=1}^N k_i^+/k_i^-,
\end{equation}
where Boltzmann's constant $k_B$ multiplied by the temperature $T$ is set to $k_BT=1$ throughout  in this paper. For example, 
if one $ATP$ is consumed and $ADP+P_i$ generated in the cycle in Eq. \eqref{fullreaction}, then the affinity 
is the chemical potential difference $\A=\mu_{ATP}-\mu_{ADP}-\mu_{P_i}$.

\begin{figure}
\includegraphics[width=65mm]{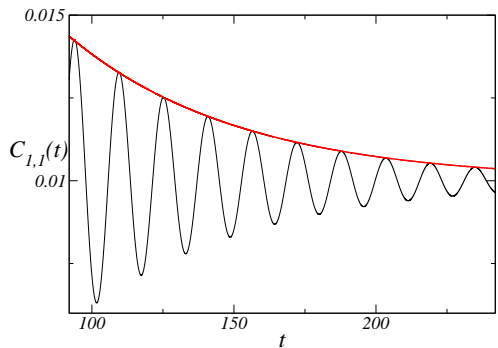}
\vspace{-2mm}
\caption{(Color online) Correlation function $C_{1,1}(t)$. 
The number of states is $N=100$, $\A=200$, and the transition rates are uniform with $k^-=1$.
For this case $X_I\simeq 0.401$, which gives a period of $15.66$, and $X_R\simeq0.01655$,
as indicated by the red solid line.
}
\label{fig1} 
\end{figure}

The model from Eq. \eqref{fullreaction} follows the master equation $d\mathbf{P}(t)/dt= \mathbf{L}\mathbf{P}(t)$, 
where $\mathbf{P}(t)=\{P_1(t),P_2(t),\ldots,P_N(t)\}^T$ is the vector of probabilities to be in a certain state.
The stochastic matrix $\mathbf{L}$
is defined by
\begin{equation}
\mathbf{L}_{j,i}\equiv k_i^{+}\delta_{i,j-1}+k_i^{-}\delta_{i,j+1}-(k_i^{-}+k_i^{+})\delta_{i,j},
\end{equation}
where $\delta_{i,j}$ is the Kronecker delta, $j-1=N$ for $j=1$, and $j+1=1$ for $j=N$. Let us assume that the enzyme is phosphorylated only in state $E_1$. The precision 
of oscillations in the phosphorylation level of an enzyme that is phosphorylated at time $t=0$ is characterized by the number of coherent oscillations in the correlation 
function $C_{1,1}(t)$ plotted in Fig. \ref{fig1}, which is the probability that the enzyme is in state $E_1$ at time $t$ given that the enzyme was in state $E_1$ at time $0$, i.e.,
\begin{equation}
C_{1,1}(t)\equiv \left[\exp(\mathbf{L}t)\mathbf{P}(0)\right]_1,
\label{eqcorr}
\end{equation}
where $\mathbf{P}(0)=\{1,0,0,\ldots,0\}$ and the subscript $1$ indicates the first component of the vector $\exp(\mathbf{L}t)\mathbf{P}(0)$.
For large $t$, this correlation function tends to $P^{st}_1$, which is the stationary distribution for state $1$. This stationary distribution is the right eigenvector of 
the stochastic matrix $\mathbf{L}$ that is associated with the eigenvalue $0$. 

The first nontrivial eigenvalue of the stochastic matrix  $\lambda= -X_R\pm X_Ii$, gives the decay time $X_R^{-1}$ and the period of oscillations $2\pi/X_I$ in Fig. \ref{fig1}. 
We characterize the coherence of oscillations by the ratio \cite{qian00}
\begin{equation}
\R\equiv X_I/X_R,
\label{eqratio}
\end{equation}
where the number of coherent oscillations \cite{more07,cao15} is $X_I/(2\pi X_R)=\R/(2\pi)$ .  

For general Markov processes that fulfill detailed balance, which corresponds to $\A=0$ for the unicyclic model, 
$X_I=0$ and there are no oscillations in correlation functions. Hence, a non-zero driving 
affinity $\A$ is a necessary condition for biochemical oscillations. In particular, for the case of uniform
rates in Eq. \eqref{fullreaction} given by $k_i^{-}=k^-$ and $k_i^{+}=k^-\textrm{e}^{\A/N}$, we obtain 
$X_R= [1-\cos(2\pi/N)](k_++k_-)$ and $X_I=\sin(2\pi/N)(k_+-k_-)$.

For the general unicyclic scheme in Eq. \eqref{fullreaction} with fixed affinity $\A$ and number of states $N$,
the ratio $\R$ is maximized for uniform transition rates, which leads to our first main result
\begin{equation}
\R\le \cot(\pi/N)\tanh[\A/(2N)]\equiv f(\A,N).
\label{firstmain}
\end{equation}
Thus, the maximal number of coherent oscillations in a unicyclic network is bounded by 
the thermodynamic force $\A$ and by the network topology through the number of states $N$.
The evidence for this bound is as follows. For $N=3$ we can show analytically that uniform rates correspond to a maximum of $\R$,
whereas for larger $N$ we rely on extensive numerical evidence as shown in Appendix \ref{app1}. Specifically, we have confirmed this conjecture
up to $N=8$ with both numerical maximization of $\R$ and evaluation of $\R$ at randomly chosen rates. Similar to the ratio $\R$, 
the Fano factor associated with the probability current is extremized for uniform rates \cite{bara15a,bara15}. However, 
as discussed in Appendix \ref{app2}, the bound in Eq. \eqref{firstmain} and this earlier bound on the Fano factor are different results, i.e., 
one does not imply the other.

%==========================================================================
\section{Multicyclic networks}
%==========================================================================
\label{sec3}

\begin{figure}
\subfigure[]{\includegraphics[width=40mm]{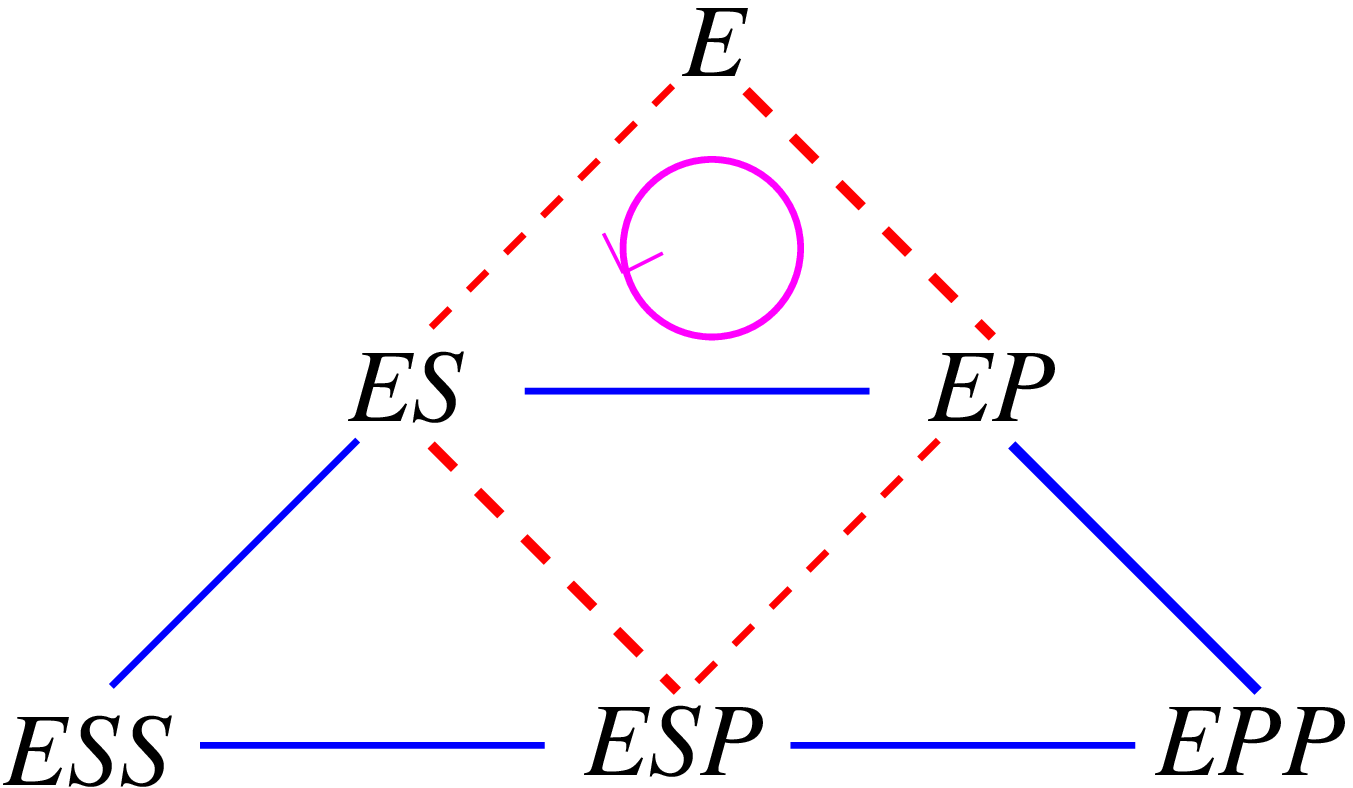}\label{fig2a}}
\subfigure[]{\includegraphics[width=40mm]{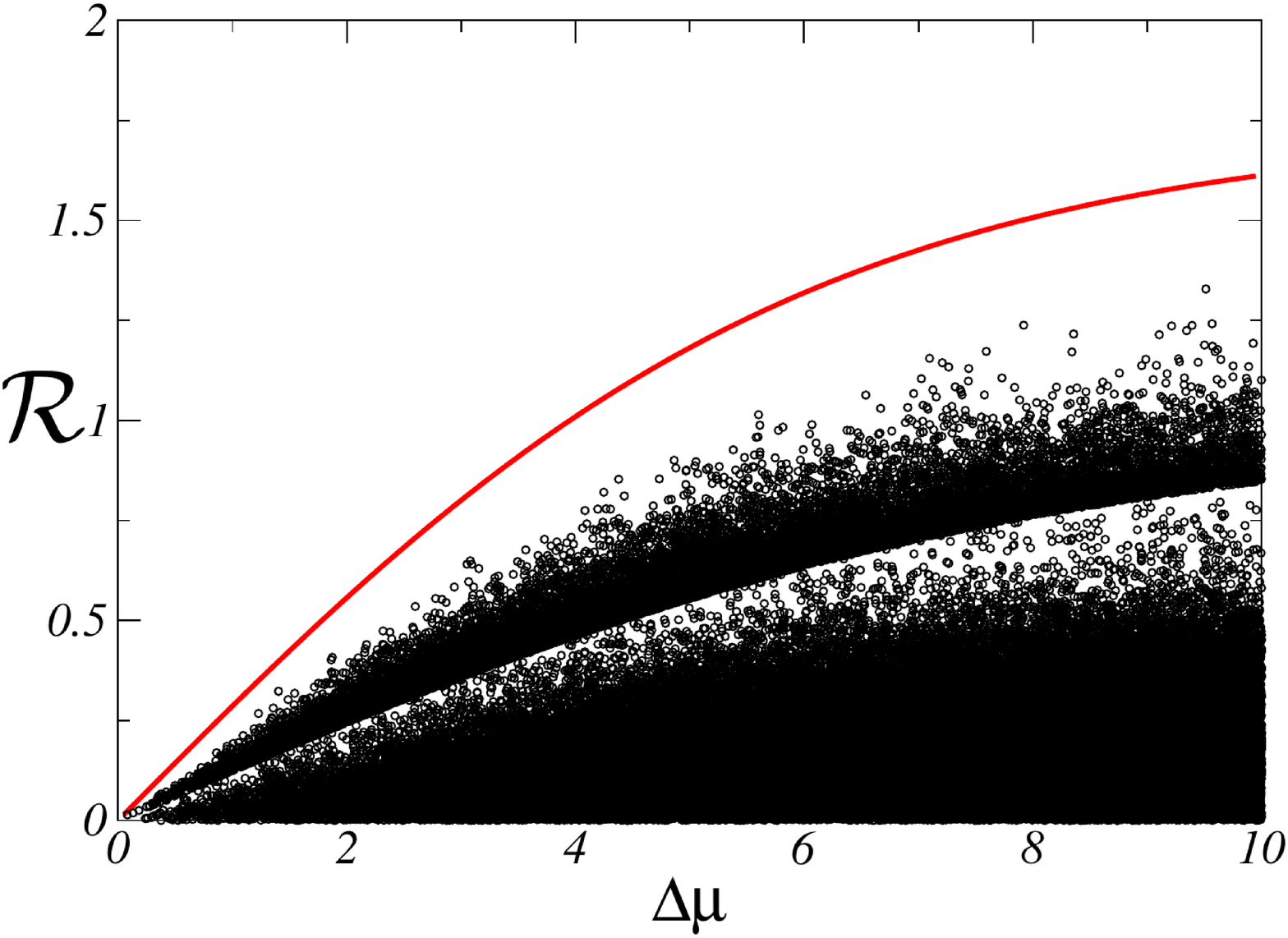}\label{fig2b}}
\vspace{-2mm}
\caption{(Color online) Multicyclic network. (a) Network of states, with the three-state cycle
marked with a magenta circle, the four-state cycle with the dashed red links and the five-state
cycle with the solid blue links. (b) Numerically evaluated $\R$ for randomly chosen rates against 
the bound (solid red line). The points were generated according to the method explained in Appendix \ref{app3}.
}
\label{fig2} 
\end{figure}

Biochemical networks are typically more complicated than a single cycle. We now extend the bound
from Eq. \eqref{firstmain} to general multicyclic networks. As an example, we consider an enzyme 
$E$ that consumes a substrate $S$ and generates a product $P$. The enzyme has two binding sites,
leading to the network of states shown in Fig. \ref{fig2a}, which is a common model in enzyme 
kinetics \cite{bara15}. The affinity that drives the system out of equilibrium is the 
chemical potential difference between substrate and product $\varDelta\mu=\mu_S-\mu_P$. 

This network of states has four types of cycles, as 
shown in Fig. \ref{fig2a}. There are cycles with three states and affinity $\varDelta\mu$, 
like the cycle $E+S\to ES\to EP\to E+P$; cycles with four states and affinity $0$, 
like the cycle $E+S+P\to ES+P\to ESP\to EP+S\to E+S+P$; cycles with five states and affinity 
$\varDelta\mu$, like the cycle $ES+S\to ESS\to ESP\to EPP\to EP+P\to ES+P$; and one cycle with six 
states and affinity $2\varDelta\mu$, which is the outer cycle in  Fig. \ref{fig2a} that goes through all 
states. Among all these cycles, the last one with $\A=2\varDelta\mu$ and $N=6$ leads to the largest value 
of the function $f(\A,N)$. We have verified numerically that indeed $f(2\varDelta\mu,6)$ bounds 
the ratio $\R$ with numerical maximization and  numerical evaluation at
randomly chosen rates, as shown in Fig. \ref{fig2b}. The bound is saturated if the transition rates for the cycle 
with six states are uniform and much faster than the rates associated with the three links in
the middle that are not part of the six-state cycle. In this way, the multicyclic network corresponds effectively 
to a unicyclic network with six states. In Appendix \ref{app3}, we perform similar numerical tests 
for several multicyclic networks that do not share any symmetry, and in all cases the ratio
$\R$ follows a similar bound.

Based on this numerical evidence we conjecture the following universal bound on the ratio $\R$. 
Consider an arbitrary Markov process with a finite number of states $N$ on an arbitrary multicyclic network. The cycles are labeled by $\alpha$,
with a number of states $N_\alpha\le N$ and affinity $\A_\alpha$, where $\textrm{e}^{\A_\alpha}$ is the product of forward transition rates divided by 
backward transition rates over all links in the cycle (see Appendix \ref{app3}). The affinity and number of states of the cycle with the maximal value of 
$f(\A_\alpha,N_\alpha)$, defined in Eq. \eqref{firstmain}, are denoted by $\A^*$ and $N^*$, respectively, i.e.,  $f(\A^*,N^*)=\textrm{max}_\alpha f(\A_\alpha,N_\alpha)$.
The ratio $\R$ is then bounded by
\begin{equation}
\R\le f(\A^*,N^*)\le \A^*/(2\pi).
\label{secondmain}
\end{equation}
The basic idea behind this bound is that the simple unicyclic network in Eq. \eqref{fullreaction} is a building block 
for a generic multicyclic network: any two point correlation function cannot 
have a larger number of coherent oscillations than the bound determined by its ``best'' cycle. Hence,
the number of coherent oscillations is bounded by the thermodynamic force $\A^*$ and the topology of the network 
of states, as characterized by $N^*$. Our bound in Eq. \eqref{secondmain} leads to two general necessary conditions 
for a large number of coherent oscillations, a large number of states and a large maximal affinity.
For biochemical models with irreversible transitions, e.g., the models 
in \cite{more07,vanz07},  the affinity $\A^*$ formally diverges, and the bound in Eq. \eqref{secondmain} becomes 
$\R\le \cot(\pi/N^*)\le \cot(\pi/N)$. The weaker second inequality involving the total number of states $N\ge N^*$ follows from 
a known result about the eigenvalues of a discrete time stochastic matrix \cite{dmit45,dmit46,swif72}. For the case of a 
complex network of states where identifying the large number of states in a cycle is not feasible, like in the activator-inhibitor 
model below, we can use the second inequality in Eq. \eqref{secondmain}, based on $\lim_{N\to\infty}f(\A,N)=\A/(2\pi)$.
We now proceed to illustrate this second main result in two models.

%==========================================================================
\section{Case studies}
%==========================================================================
\label{sec4}

%==========================================================================
\subsection{Model for a single KaiC}
%==========================================================================
\label{sec4a}

\begin{figure}
\subfigure[]{\includegraphics[width=12mm]{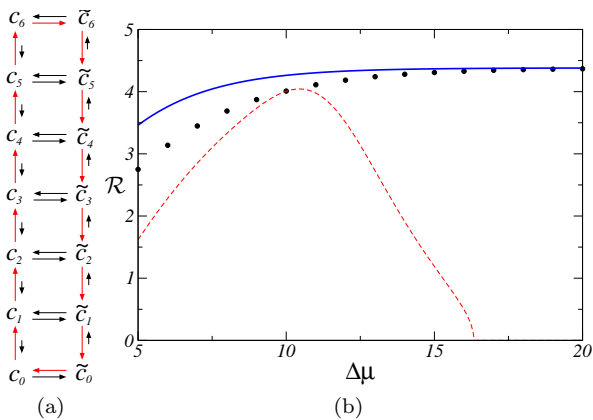}\label{fig3a}} 
\subfigure[]{\psfrag{R}{$\mathcal{R}$}\includegraphics[width=65mm]{./fig3b.eps}\label{fig3b}} 
\vspace{-2mm}
\caption{(Color online) Model and results for a single KaiC protein. (a) For the vertical arrows, the transition rates 
are $\gamma\textrm{e}^{\varDelta\mu/2}$ for the larger arrow and $\gamma\textrm{e}^{E/6}$ for the small arrows.
For the horizontal arrows, the transition rate from $C_i$ to $\tilde{C}_i$ is $k\textrm{e}^{\chi E(i-3)/3}$ and 
the transition rate from $\tilde{C}_i$ to $C_i$ is $k\textrm{e}^{\tilde{\chi}E(3-i)/3}$, where $\chi$ ($\tilde{\chi}$)
is an indicator function that is zero (one) for $i=0,1,2,3$ and one (zero) for $i=4,5,6$. (b) Ratio $\R$ as a function of 
$\varDelta \mu$. The dots were obtained 
by numerical maximization of $\R$ with respect to the parameters $\gamma$ and $E$, where $k=1$. The dotted red line represents $\R$ for
$k=1$, $\gamma=\textrm{e}^5$, and $E=10$. The blue solid line is the bound $f(\A,N)$ for $\A=6 \varDelta\mu$ and $N=14$. 
}
\label{fig3} 
\end{figure}

First, we consider a model for a single KaiC hexamer along the lines of the  model proposed in \cite{vanz07}. 
The assumptions entering the model, which is depicted in Fig. \ref{fig3a}, are the following.
A phosphate can bind to each one of the six monomers, hence the phosphorylation level of the hexamer varies from $i=0$, with 
no phosphate, to $i=6$, with all monomers phosphorylated. Each of the six monomers can be either active or inactive. However, either
all monomers are active or all monomers are inactive, since the energetic cost of having two monomers with different conformations 
is high enough to avoid such configurations. There are a total of $14$ states, denoted by $C_i$ for $i$ phosphorylated active monomers and 
$\tilde{C}_i$ for $i$ phosphorylated inactive monomers. If the hexamer is active, phosphorylation reactions occur and if 
the hexamer is inactive only dephosphorylation reactions occur.

The transition rates of this model for a single KaiC protein are given in the caption of Fig. \ref{fig3a}. The parameter $\varDelta \mu$
is the chemical potential difference of ATP hydrolysis. The parameter $E$ sets the energy of a state that depends on the hexamer
activity and on the phosphorylation level. If the hexamer is active the energy of a state $C_i$ is $Ei/6$ and if it is
inactive the energy of an state $\tilde{C}_i$ is $E(6-i)/6$. This parametrization implies that the transition rate from $C_6$ to $\tilde{C}_6$ and 
the transition rate from $\tilde{C}_0$ to $C_0$ are both larger than the rates for the respective reversed transitions. The parameters $k$ and $\gamma$ are related 
the time-scales of changes in the phosphorylation level and conformational changes between active and inactive, respectively.

The phosphorylation level of the KaiC protein oscillates with the number of coherent oscillations given by $\R/(2\pi)$, as shown in in Appendix \ref{app4}. 
The cycle with the largest value of the function $f(\A,N)$ is the cycle that goes through all $N=14$ states with $\A=6\varDelta \mu$, which is 
marked with the red arrows in Fig. \ref{fig3a}. Hence, for this model we have $\R\le f(6\varDelta\mu,14)$, as shown in Fig. \ref{fig3b}.

For fixed $E$ and $\gamma$  we obtain the red dashed curve in Fig. \ref{fig3b} for $\R$ as a function of $\varDelta\mu$. Interestingly, while 
the number of coherent oscillations has a maximum, after which it decreases to zero with increasing $\varDelta \mu$, the entropy
production from stochastic thermodynamics \cite{seif12} is an increasing function of $\varDelta\mu$. Hence, the number of 
coherent oscillations can also decrease with an increase of the rate of free energy consumption, which provides a counter example 
to the relation between the number of coherent oscillations and energy dissipation inferred in \cite{cao15}. We discuss the relation 
between $\R$ and the entropy production further in Appendix \ref{app5}. 

The maximal number of coherent oscillations $\R/(2\pi)$ that can 
be achieved in this model is strictly speaking less than $1$. If a single molecule does not have a large number of states, 
several coherent oscillations can only be sustained in a system with many molecules as we discuss next in our second example.

%==========================================================================
\subsection{Activator-inhibitor model}
%==========================================================================
\label{sec4b}

\begin{figure}
\subfigure[]{\includegraphics[width=20mm]{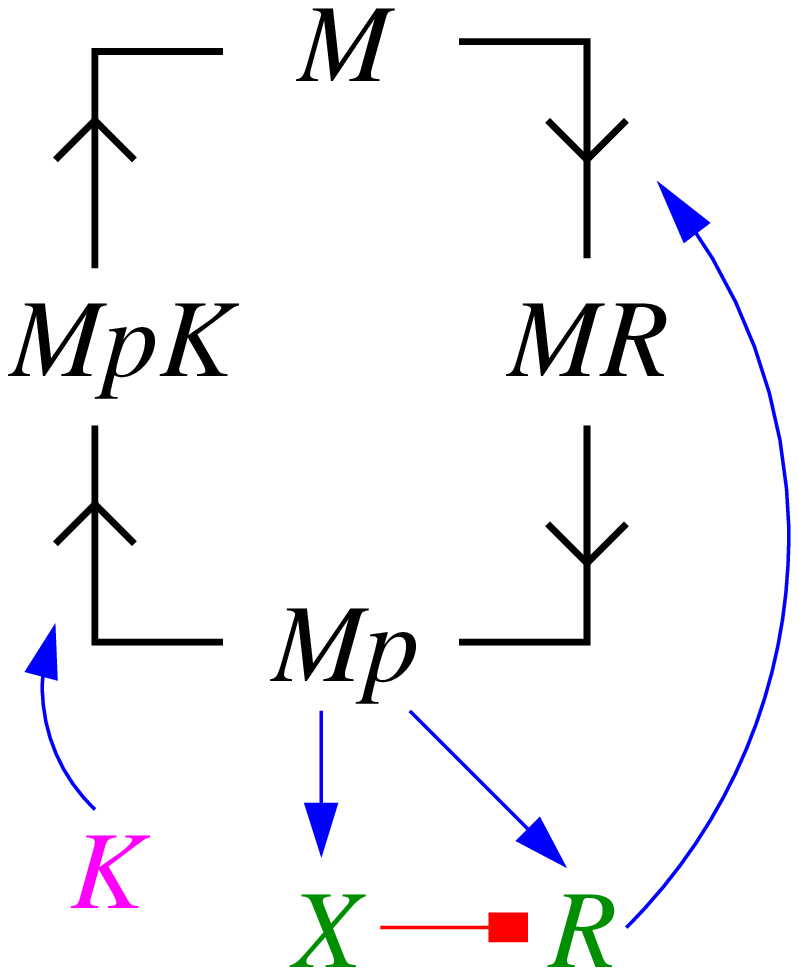}\label{fig4a}}
\subfigure[]{\psfrag{R}{$\mathcal{R}$}\includegraphics[width=65mm]{./fig4b.eps}\label{fig4b}}
\vspace{-2mm}
\caption{(Color online) The activator-inhibitor model. (a) Scheme representing the chemical reactions in the model, 
where a blue arrow represents activation and the red lines with an square at the end represents inhibition. The four
forms of the enzyme are free enzyme $M$, the species $R$ bound to the enzyme $MR$, phosphorylated enzyme $M_p$ and 
phosphorylated enzyme bound to a phosphatase $M_pK$. (b) Numerical results for the ratio $\R$. The number of enzymes is $N_M=500$.
The points were obtained from numerical simulation as explained in Appendix \ref{app6} and the solid lines represent 
the estimate of the $\R$ given by $N_K\varDelta\mu/(2\pi)$. The optimal 
number of oscillations takes place  close to $N_K=30$, with $N_K=20$ there were practically no 
oscillations and with $N_K=50$ the number of oscillations gets considerably smaller. 
}
\label{fig4} 
\end{figure}

We consider the activator-inhibitor model from \cite{cao15}, see Fig. \ref{fig4a}. The main components of this model are inhibitors $X$, activators $R$ and 
enzymes $M$ that can be in four different states. The enzyme goes through a phosphorylation cycle over these four states, hydrolyzing one ATP thus liberating a 
free energy $\varDelta\mu$. The enzyme $M$ in its phosphorylated form ($M_p$) activates $R$ and $X$, whereas $X$ inhibits $R$. Furthermore,  the enzyme $M$ must 
bind an $R$ in order to phosphorylate. Hence, $R$ activates the production of $R$ and $X$, while $X$ inhibits $R$. This feedback loop leads to oscillations 
in, for example, the number of species $X$. Finally, there is a phosphatase $K$ that must bind to the enzyme $M$ for the dephosphorylation reaction.
Further details of the model are given in Appendix \ref{app6}. 

Two important aspects about the behavior of this model are the following. First, the number of oscillations increases 
with $\varDelta \mu$ and saturates for large enough $\varDelta\mu$. Second, in order for different enzymes $M$ to synchronize their cycles,
they must compete for the smaller number of phosphatase $N_K<N_{M_T}$, where $N_{M_T}$ is the total number of enzymes, as explained in Appendix \ref{app6}. 
However, if $N_K$ is too small, the number of enzymes that 
synchronize is also too small to generate oscillations. Hence, there is an optimal value 
for $N_K$. These features are shown Fig. \ref{fig4b}.

Due to the complex network of states of this model we bound the ratio $\R$ with the second inequality in Eq. \eqref{secondmain}. 
The largest affinity $\A^*$ is given by $\A^*=\varDelta\mu N_{M_T}$, which corresponds to a cycle where all enzymes $M$ go through
their cycles in a synchronized way. 

As shown in Fig. \ref{fig4b}, the values of $\R$ obtained with numerical simulations are  approximately one order of 
magnitude below the fundamental limit set by our bound of $N_{M_T}\varDelta\mu/(2\pi)$, which gives $\R\simeq796$ for $\varDelta\mu=10$. 
This result is reasonable, as saturating the bound in a multicyclic network requires transition rates such that an optimal cycle dominates, 
which is not the case for the present model. The realization of this optimal cycle in a stochastic trajectory  
would require an unlikely sequence of events that all enzymes $M$ go through their own cycles in a 
synchronized way.

For the case of a close to optimal value of $N_K$ that maximizes $\R$, the number of enzymes $M$ that synchronize is roughly  $N_K$. Hence,
cycles with an affinity $\varDelta\mu N_K$ should be typical. Guided by our bound, it is then interesting to compare the value 
of $\R$ with the estimate $\varDelta\mu N_K/(2\pi)$. As shown in Fig. \ref{fig4b}, indeed this number can give an approximate value of $\R$,
typically overestimating $\R$. The estimate is best for $N_K=30$ and for $\varDelta\mu=10$. For large  $\varDelta\mu$ the bound grows 
linearly, while $\R$ saturates. Our results then indicate that this estimate works well in the regime where $\varDelta\mu$ is close to its saturation value and 
$N_K$ is close to its optimal value. Even though this heuristic argument is restricted to this model, if competition for some scarce 
molecule that lead to synchronization is present in the biochemical system, e.g., in the model for the Kai system from \cite{vanz07}, 
a similar reasoning that leads to an estimate of the number of coherent oscillations should be valid. 

%==========================================================================
\section{Conclusion}
%==========================================================================
\label{sec5}

In summary, we have conjectured a new bound on the number of coherent biochemical oscillations 
for systems with large fluctuations. Our universal result depends only on the thermodynamic forces 
that drive the system out equilibrium and on the network topology through the cycle with the largest 
function $f$ from Eq. \eqref{firstmain}. Knowledge of the chemical potential differences and the network 
of states of a biochemical system thus leads to a bound on the number of coherent oscillations. As illustrative 
examples, we obtained the largest number of oscillations that can be sustained by a single KaiC hexamer 
and analyzed the activator-inhibitor model, showing that our bound is also valid in models with a number of molecules
that is large enough to make the network of states complicated but small enough to keep fluctuations relevant and, therefore,
make a description in terms of deterministic rate equations inappropriate.   

It remains to be seen whether and how  our bound can be used as a guiding principle to understand how systems like 
circadian clocks have evolved and to engineer systems with precise oscillations in synthetic biology. 
Our results apply to autonomous biochemical oscillators. Analyzing the relation between precision and thermodynamics 
for biochemical oscillators that are coupled to an external periodic signal is an interesting direction for future work.
Finally, a rigorous mathematical proof of our conjecture about the first excited eigenvalue of a stochastic matrix is
an open problem for the theory of Markov processes.

\begin{acknowledgments} 
We thank F. J\"ulicher for helpful discussions.
\end{acknowledgments}

%==========================================================================
\appendix
%==========================================================================

%==========================================================================
\section{Evidence for unicyclic network}
%==========================================================================
\label{app1}

We discuss the evidence for the bound in Eq. \eqref{firstmain} for the unicyclic scheme. The  mathematical problem 
is to calculate the first non-trivial eigenvalue of the stochastic matrix
\begin{equation}
\mathbf{L}_{j,i}\equiv k_i^{+}\delta_{i,j-1}+k_i^{-}\delta_{i,j+1}-(k_i^{-}+k_i^{+})\delta_{i,j},
\end{equation}
where $\delta_{i,j}$ is the Kronecker delta, $j-1=N$ for $j=1$, and $j+1=1$ for $j=N$. The absolute value of the imaginary (real)
part of this eigenvalue is denoted $X_I$ ($X_R$).

For the case $N=3$ this eigenvalue can be exactly calculated with some algebra, leading to 
\begin{equation}
\R\equiv X_I/X_R= \sqrt{4C_1/C_2^2-1},
\end{equation}
where 
\begin{align}
C_1\equiv & k_1^-k_2^-+ k_1^-k_3^-+ k_2^-k_3^-+k_1^+k_2^++ k_1^+k_3^++ k_2^+k_3^+\nonumber\\
& +k_1^-k_2^++k_2^-k_3^++k_3^-k_1^+
\end{align}
and
\begin{equation}
C_2\equiv k_1^++k_2^++k_3^++k_1^-+k_2^-+k_3^-. 
\end{equation}
If $4C_1<C_2^2$, there are no oscillations in correlations functions. We want to find the transition rates that maximize $\R$ for a fixed affinity 
\begin{equation}
\A\equiv \ln\left(\prod_{i=1}^N k_i^+/k_i^-\right).
\end{equation}
This maximum can be found with the Lagrange function
\begin{equation}
\Lambda(\{k_i^+\},\{k_i^-\},\alpha)= 4C_1/C_2^2-\alpha(k_1^+k_2^+k_3^+-k_1^-k_2^-k_3^-\textrm{e}^\A),
\end{equation}
where $\alpha$ is a Lagrange multiplier. The derivatives of $\Lambda$ with respect to the transition rates are given by 
\begin{equation}
\frac{d\Lambda}{dk_1^+}=\frac{4(k_2^++k_3^++k_3^-)}{C_2^2}-8\frac{C_1}{C_2^3}-k_2^+k_3^+\alpha
\end{equation}
and
\begin{equation}
\frac{d\Lambda}{dk_1^-}=\frac{4(k_2^-+k_2^++k_3^-)}{C_2^2}-8\frac{C_1}{C_2^3}+\textrm{e}^\A k_2^-k_3^-\alpha.
\end{equation}
Due to symmetry, it is easy to deduce the derivatives with respect to $k_2^\pm$ and $k_3^\pm$ from the above expressions.
If we substitute uniform rates $k_i^-=k^-$ and $k_i^+= \textrm{e}^{\A/3}k_-$ in the above expressions, we obtain that both derivatives
become zero with a Lagrange multiplier
\begin{equation}
\alpha=\frac{4\textrm{e}^{-2\A/3}(\textrm{e}^{\A/3}-1)}{9(\textrm{e}^{\A/3}+1)^3(k^-)^3}.
\end{equation}
Hence, we have proved that $\R$ is extremized for uniform rates. We can easily evaluate $\R$ for specific rates 
and check that uniform rates indeed correspond to  a maximum.

\begin{figure}
\includegraphics[width=75mm]{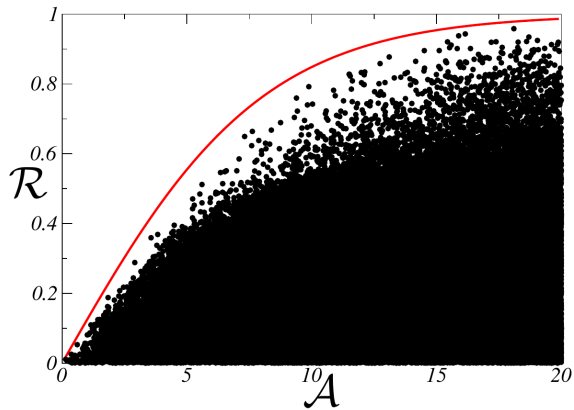}
\vspace{-2mm}
\caption{(Color online) Scatter plot for the unicyclic model with $N=4$. The bound in Eq. \eqref{firstmain} is represented by the solid red line. 
The rate $k_1^+$ was set to $k_1^+= \textrm{e}^{A}k_1^-k_2^-k_3^-/(k_2^+k_3^+)$,
the other seven rates were randomly chosen as $10^{x}$, with $x$ uniformly distributed between $-3$ and $3$. For this figure, we have 
evaluated $\R$ for $10^{7}$ sets of rates.}
\label{figsup1} 
\end{figure}

For larger $N$, up to $N=8$, we have calculated this eigenvalue numerically. We have maximized the ratio $\R$ numerically and 
observed that it is maximized for uniform rates in all cases, providing  convincing evidence for the bound in Eq. \eqref{firstmain}. 
As an independent check we have also evaluated $\R$ numerically for randomly chosen rates. As an example, we show a scatter plot obtained 
with this method in Fig. \ref{figsup1}.

%==========================================================================
\section{Relation between $\R$ and the Fano factor}
%==========================================================================
\label{app2}

We now explain the difference between the bound in Eq. \eqref{firstmain}  and a bound on the Fano factor $F$ obtained in \cite{bara15a,bara15}.
This Fano factor is given by $F=2D/J$, where $J$ is the average probability current and $D$ the diffusion constant associated with the current \cite{bara15a}.
This bound on the Fano factor can be written as $F^{-1}\le N\tanh[\A/(2N)]$, where the quantity $F^{-1}$ is also maximized for uniform rates.
Furthermore, for uniform rates $2D=(k_++k_-)/N^2$ and  $J=(k_+-k_-)/N$, implying $X_I= \sin(2\pi/N)NJ$ and $X_R= [1-\cos(2\pi/N)]N^22D$. Nevertheless,
for arbitrary transition rates there is no such simple relations, with a prefactor that only depends on $N$, between $X_I$ ($X_R$) and $J$ ($D$). 

For uniform rates $\R= F^{-1}N^{-1}\cot(\pi/N)$. Since $\R$ can be zero even out of equilibrium and $F^{-1}$ becomes 
zero only in equilibrium, we know that $F^{-1}N^{-1}\cot(\pi/N)$ can be larger than $\R$. Evaluating $\R$ and $F$ at different rates 
we find that $\R$ can also be larger than $F^{-1}N^{-1}\cot(\pi/N)$. Hence, the bound on the Fano factor from \cite{bara15a,bara15}
does not imply our result in Eq. \eqref{firstmain}. Their main similarity is that both the Fano factor $F$ and the ratio $\R$ are extremized for uniform rates:
it is common to find a function of several variables that is extremized at a symmetric point.

%==========================================================================
\section{Evidence for multicyclic networks}
%==========================================================================
\label{app3}

In this appendix we explain the numerical evidence for the bound in Eq. \eqref{secondmain} for multicyclic networks.  For all cases,
we have confirmed our bound with numerical calculation of the first nontrivial eigenvalue of the stochastic matrix. We have confirmed
the bound with both numerical maximization of $\R$ and by evaluating $\R$ at randomly chosen transition rates.

\begin{figure}
\subfigure[]{\includegraphics[width=25mm]{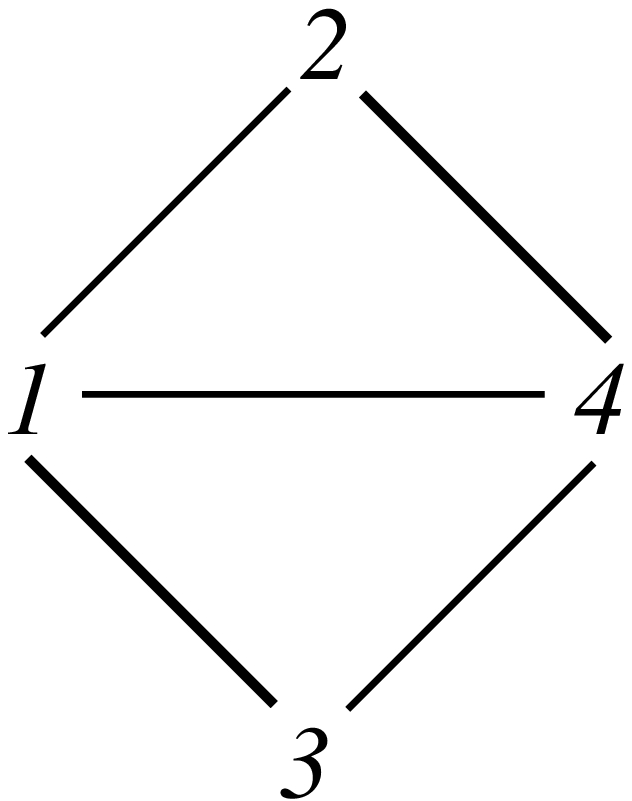}\label{figmulti1a}}
\subfigure[]{\includegraphics[width=40mm]{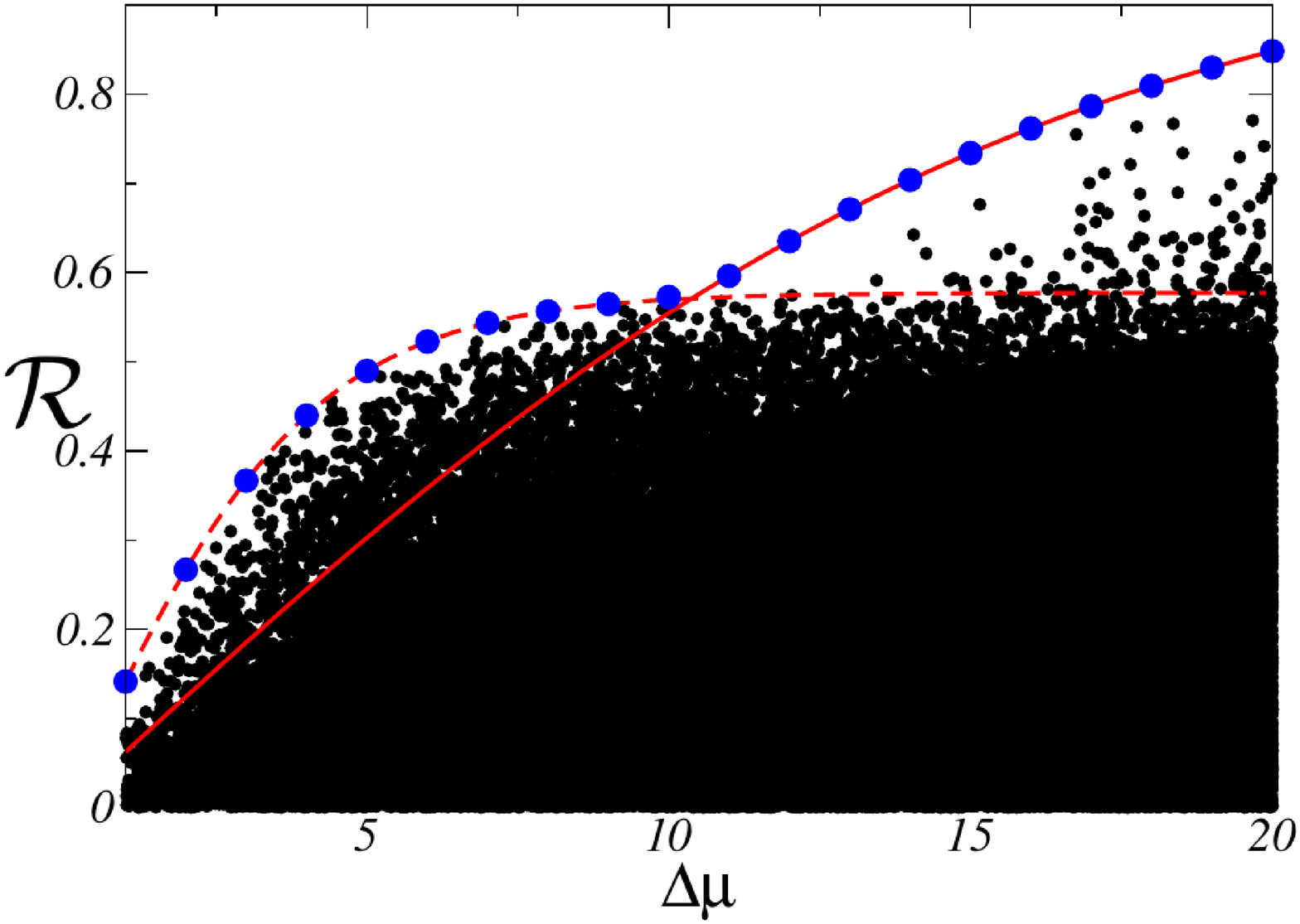}\label{figmulti1b}}
\vspace{-2mm}
\caption{(Color online) First example of a multicyclic network. (a) Network of states. (b) Numerical evidence of the bound.  
The dashed red line is the function $f(3\Delta\mu/2,3)$ and the solid red line the function $f(\Delta\mu/2,4)$.
The large blue dots were obtained with numerical maximization of $\R$ and the black dots were obtained by evaluating
$R$ at randomly generated transition rates. We have generated two sets of $10^7$ points each. For one set  we have chosen eight independent rates as: 
$k_{24}$, $k_{43}$, and $k_{31}$ were set to $\textrm{e}^{\varDelta \mu/8}10^x$; $k_{21}$, $k_{34}$, and $k_{42}$ were set to $10^x$ ; 
$k_{14}$ was set to $\textrm{e}^{-5\varDelta \mu/4}10^{-5}10^x$ and $k_{41}$ as set to $10^{-5}10^x$; where $x$
is uniformly distributed between $-3$ and $3$. The remaining rates $k_{12}$ and  $k_{13}$ were determined by the relations \eqref{eqaff1} and \eqref{eqaff2}, respectively.
For the other set the  8 independent rates were chosen as $10^x$.  
}   
\label{figmulti1} 
\end{figure}

As a first example of a multicyclic network, we consider the network with four states shown in Fig. \ref{figmulti1a}. The numbers 
represent states and the links between them represent transition rates that are not zero. A transition rate from state $i$ to $j$ is denoted $k_{ij}$.
This network has three cycles: two cycles  with three states $\mathcal{C}_1=(1,2,4,1)$ and $\mathcal{C}_2=(1,3,4,1)$, 
and one cycle with four states  $\mathcal{C}_3=(1,2,3,4,1)$.  The affinity of cycle  $\mathcal{C}_1$ is
\begin{equation}
\A_1\equiv\ln\frac{k_{12}k_{24}k_{41}}{k_{21}k_{42}k_{14}},
\label{eqaff1}
\end{equation}
 and the affinity of cycle  $\mathcal{C}_2$ is
 \begin{equation}
\A_2\equiv\ln\frac{k_{13}k_{34}k_{41}}{k_{31}k_{43}k_{14}}.
\label{eqaff2}
\end{equation}
The affinity of $\mathcal{C}_3$, which is not independent of  $\A_1$ and $\A_2$, can be written as 
\begin{equation}
\A_3=\A_1-\A_2.
\end{equation}

For the results shown in Fig. \ref{figmulti1b}, which confirm our bound for this network, we have set the affinities of the cycles as $\A_1= 3 \varDelta\mu/2$ and $\A_2= \varDelta\mu$, which leads 
to $\A_3=\varDelta\mu/2$. As shown in Fig. \ref{figmulti1b}, the cycle with the largest value for the function $f(\A_\alpha,N_\alpha)$, where $N_\alpha$ is 
the number of states of cycle $\alpha$, depends on the value of $\varDelta\mu$. For $\varDelta\mu<10.38$ the cycle with largest value for this function is 
$\mathcal{C}_1$  with $f(3\varDelta\mu/2,3)$. For $\varDelta\mu>10.38$ the cycle with largest value for this function is 
$\mathcal{C}_3$  with $f(\varDelta\mu/2,4)$.

\begin{figure}
\subfigure[]{\includegraphics[width=25mm]{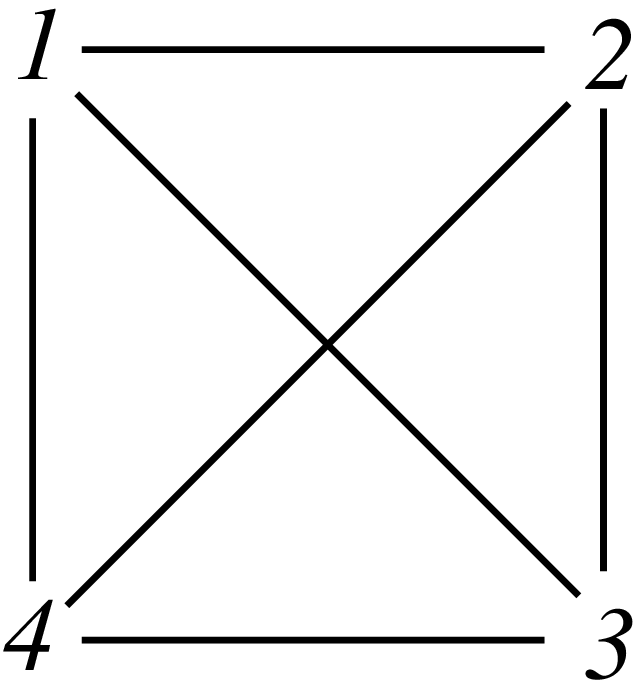}\label{figmulti2a}}
\subfigure[]{\includegraphics[width=40mm]{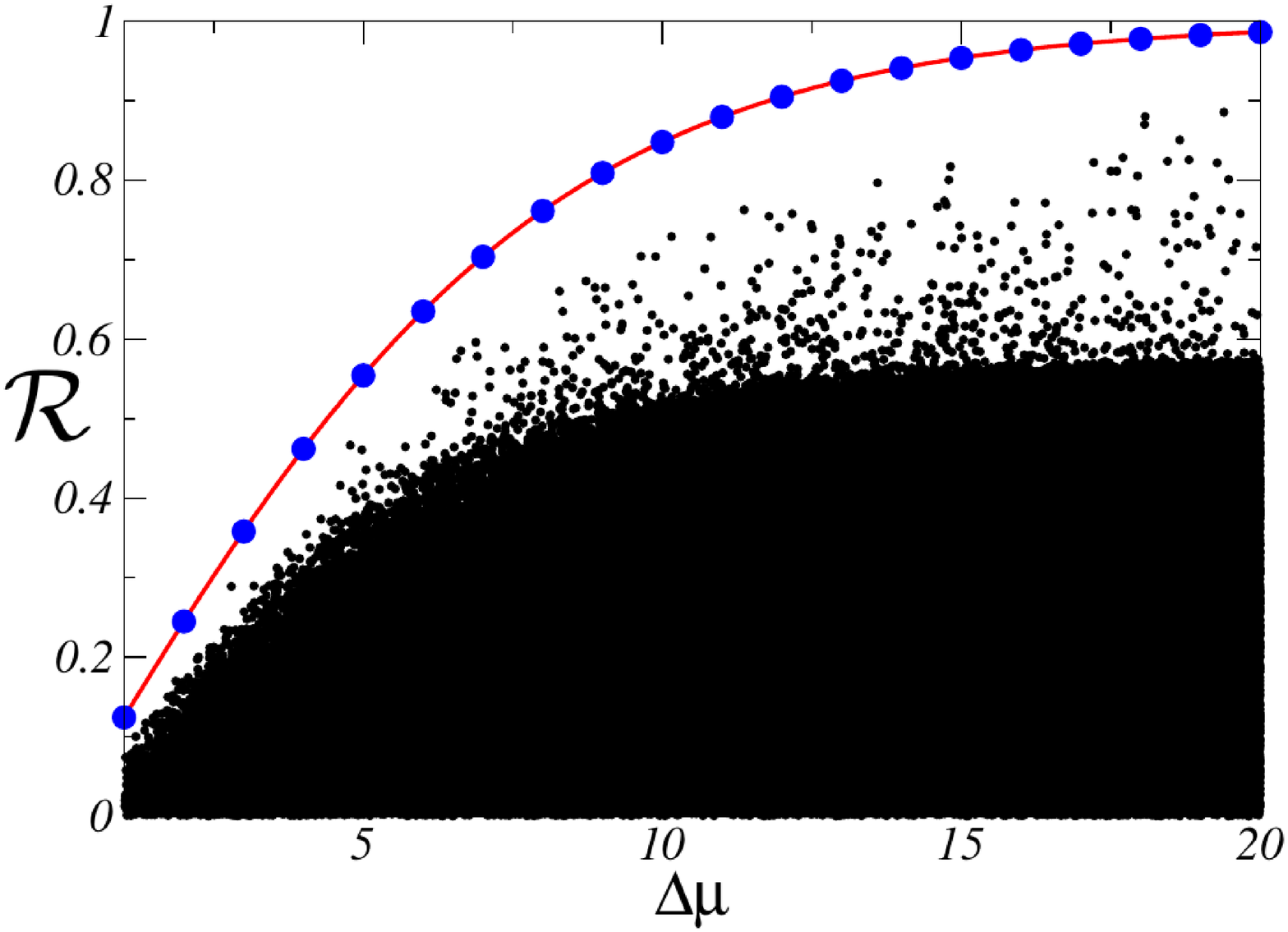}\label{figmulti2b}}
\vspace{-2mm}
\caption{(Color online) Second example of a multicyclic network. (a) Network of states. (b) Numerical evidence of the bound.  
The solid red line is the function $f(\Delta\mu,4)$. The large blue dots were obtained with numerical maximization of $\R$ and the black dots were obtained by evaluating
$R$ at randomly generated transition rates. We have generated $10^7$ points, choosing 9 independent transition rates as 
$10^x$, with $x$ uniformly distributed between $-2$ and $2$. The remaining transition rates $k_{12}$,  $k_{13}$, and $k_{42}$ were determined by the affinities in 
Eqs.  \eqref{eqaff3} ,  \eqref{eqaff4} , and   \eqref{eqaff5}.}   
\label{figmulti2} 
\end{figure}

The second multicyclic network has four states and is fully connected, as shown in Fig. \ref{figmulti2a}. For this network we have one cycle four states and four 
cycles with three states. We fix the affinities of these cycles as 
\begin{equation}
\A_1= \ln\frac{k_{12}k_{23}k_{34}k_{41}}{k_{21}k_{32}k_{43}k_{14}}=\varDelta\mu,
\label{eqaff3}
\end{equation}
\begin{equation}
\A_2= \ln\frac{k_{12}k_{23}k_{31}}{k_{21}k_{32}k_{13}}=0,
\label{eqaff4}
\end{equation}
\begin{equation}
\A_3= \ln\frac{k_{12}k_{24}k_{41}}{k_{21}k_{42}k_{14}}=0,
\label{eqaff5}
\end{equation}
where from now on we define the cycles through their affinities.
These three affinities determine the values of the two remaining affinities as
\begin{equation}
\A_4= \ln\frac{k_{13}k_{34}k_{41}}{k_{31}k_{43}k_{14}}=\varDelta\mu,
\end{equation}
\begin{equation}
\A_5= \ln\frac{k_{23}k_{34}k_{42}}{k_{32}k_{43}k_{24}}=\varDelta\mu.
\end{equation}
The cycle leading to the maximal value of $f(\A_\alpha,N_\alpha)$ is the cycle with four states and affinity $\A_1=\varDelta\mu$. 
The numerical results illustrating the bound $\R\le f(\varDelta\mu,4)$ for this network is shown in Fig. \ref{figmulti2b}.

\begin{figure}
\subfigure[]{\includegraphics[width=40mm]{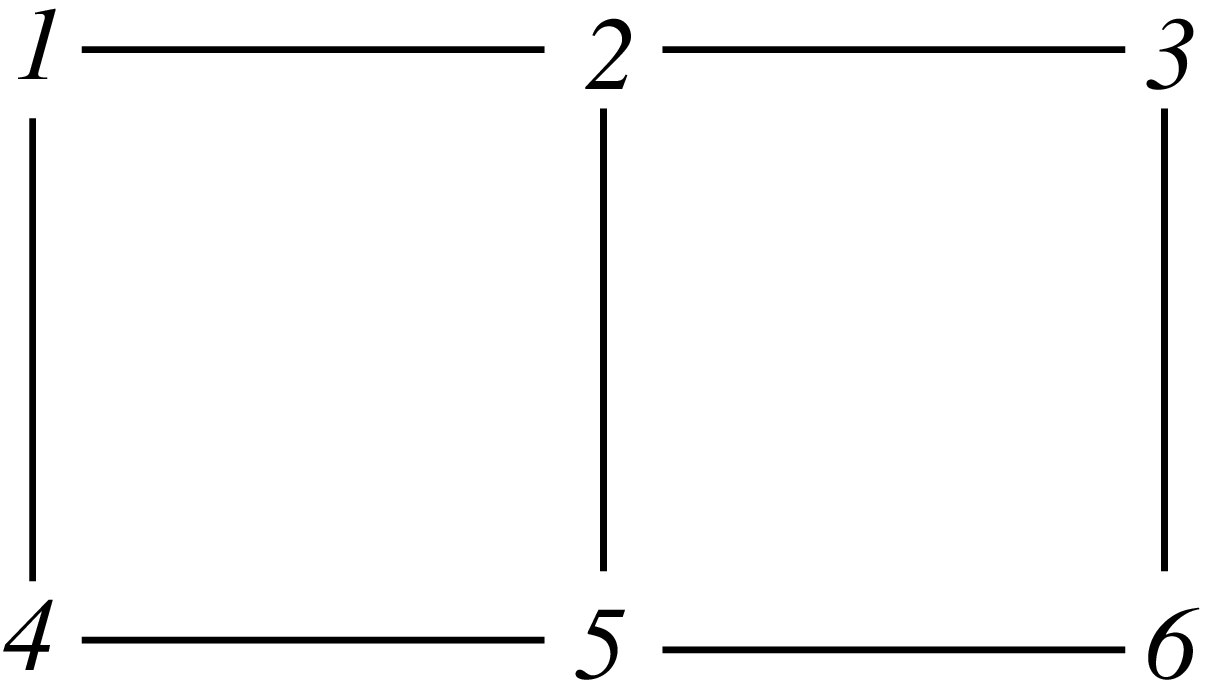}\label{figmulti3a}}
\subfigure[]{\includegraphics[width=40mm]{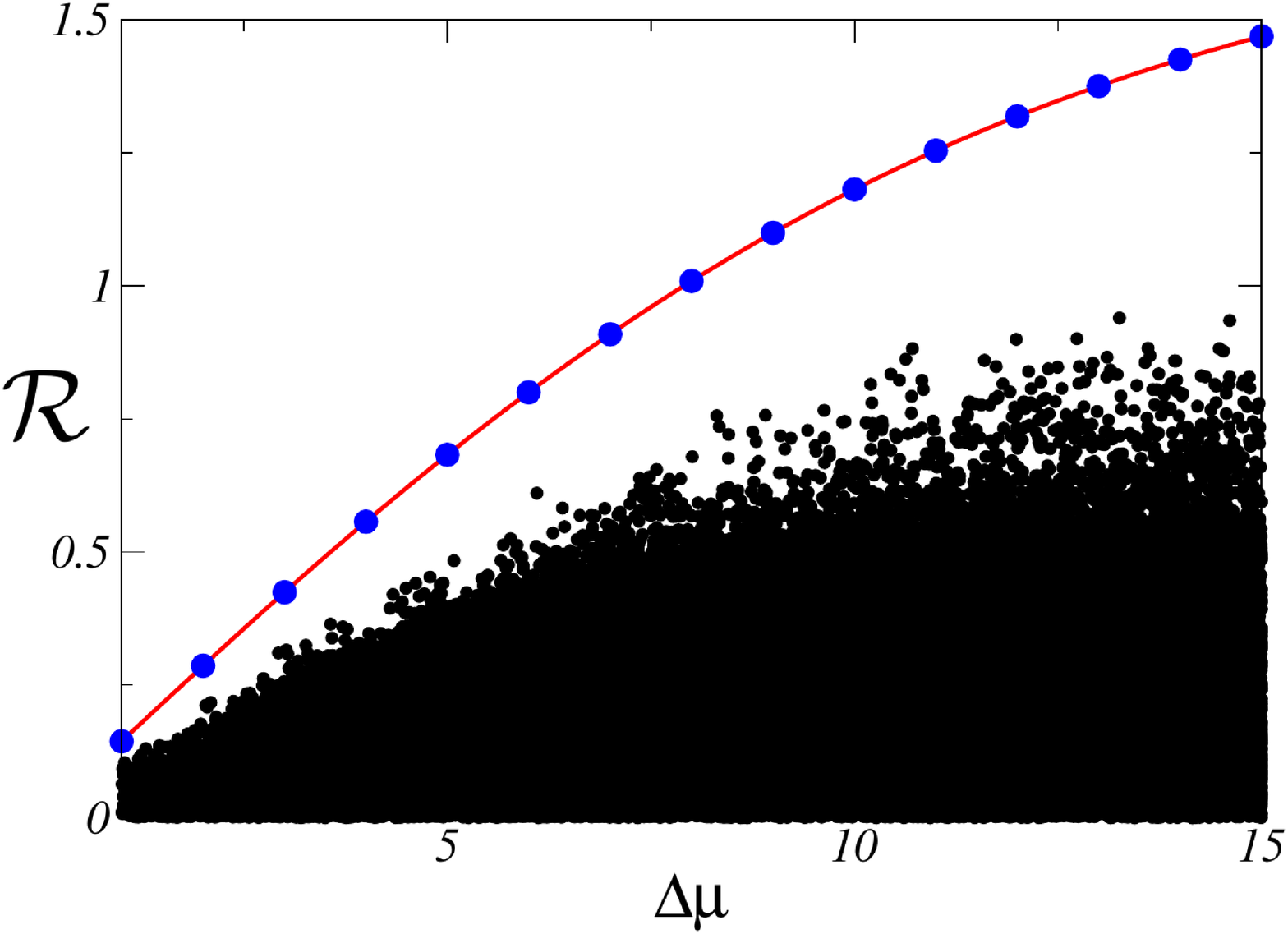}\label{figmulti3b}}
\vspace{-2mm}
\caption{(Color online) Third example of a multicyclic network. (a) Network of states. (b) Numerical evidence of the bound.  
The solid red line is the function $f(\Delta\mu,6)$. The large blue dots were obtained with numerical maximization of $\R$ and the black dots were obtained by evaluating
$R$ at randomly generated transition rates. We have generated $10^7$ points, choosing 12 independent transition rates as 
$10^x$, with $x$ uniformly distributed between $-2$ and $2$. The two remaining transition, $k_{25}$  and $k_{36}$, were determined by the affinities in 
Eqs.  \eqref{eqaff1multi3}  and  \eqref{eqaff2multi3}.}   
\label{figmulti3} 
\end{figure}

Our third example is the network with six states and three cycles shown in Fig. \ref{figmulti3a}. The affinity of the cycle with six states is fixed as 
\begin{equation}
\A_1=\ln\frac{k_{12}k_{23}k_{36}k_{65}k_{54}k_{41}}{k_{21}k_{32}k_{63}k_{56}k_{45}k_{14}}=\varDelta\mu.
 \label{eqaff1multi3} 
\end{equation}  
We also fix the affinity 
\begin{equation}
\A_2=\ln\frac{k_{12}k_{25}k_{54}k_{41}}{k_{21}k_{52}k_{45}k_{14}}=\varDelta\mu.
 \label{eqaff2multi3} 
\end{equation}  
These two affinities determine the affinity of the third cycle as
\begin{equation}
\A_3=\ln\frac{k_{23}k_{36}k_{65}k_{52}}{k_{32}k_{63}k_{56}k_{25}}=0.
\end{equation}  
The dominant cycle is the one with affinity $\A_1=\varDelta \mu$ and six states. The numerical evidence for the bound $\R\le f(\varDelta \mu,6)$ is shown if Fig. \ref{figmulti3b}

\begin{figure}
\subfigure[]{\includegraphics[width=40mm]{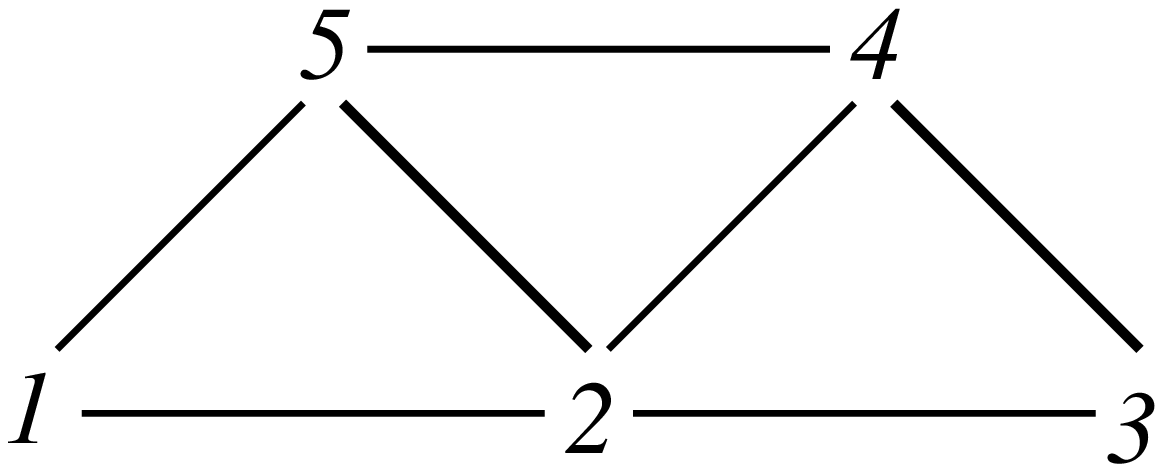}\label{figmulti4a}}\hfill
\subfigure[]{\includegraphics[width=40mm]{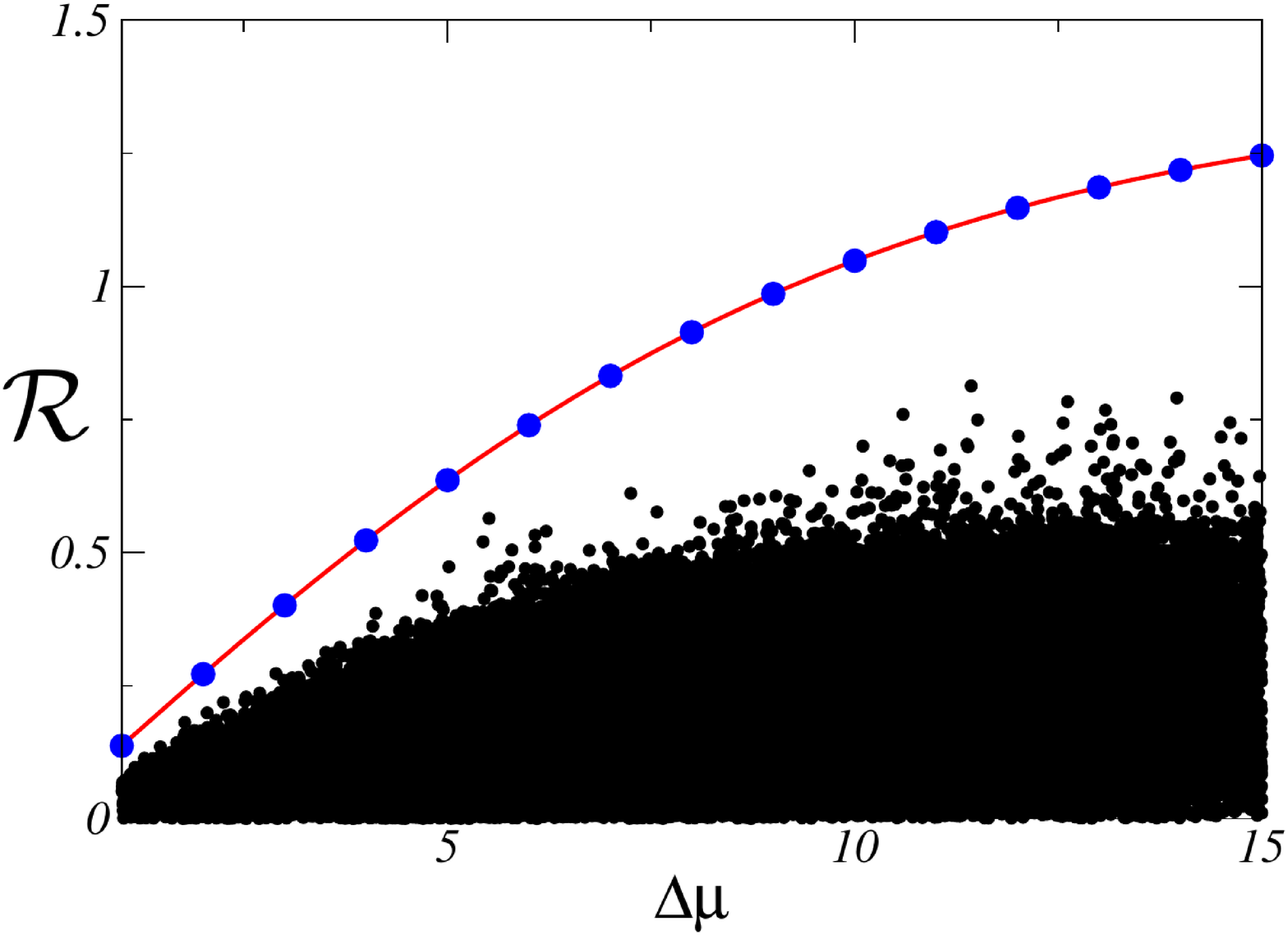}\label{figmulti4b}}\hfill
\vspace{-2mm}
\caption{(Color online) Fourth example of a multicyclic network. (a) Network of states. (b) Numerical evidence of the bound.  
The solid red line is the function $f(\Delta\mu,5)$. The large blue dots were obtained with numerical maximization of $\R$ and the black dots were obtained by evaluating
$R$ at randomly generated transition rates. We have generated $10^7$ points, choosing 11 independent transition rates as 
$10^x$, with $x$ uniformly distributed between $-2$ and $2$. The three remaining transition rates, $k_{51}$, $k_{52}$ and $k_{42}$, were determined by the affinities in 
Eqs.  \eqref{eqaff1multi4},  \eqref{eqaff2multi4} and \eqref{eqaff3multi4}.}   
\label{figmulti4} 
\end{figure}

The fourth and last example is the network with five states and five cycles shown in  Fig. \ref{figmulti4a}. 
The affinity of the five states cycle is set to 
\begin{equation}
\A_1=\ln\frac{k_{12}k_{23}k_{34}k_{45}k_{51}}{k_{21}k_{32}k_{43}k_{54}k_{15}}=\varDelta\mu.
 \label{eqaff1multi4} 
\end{equation}
There are two four states cycles with affinities
\begin{equation}
\A_2=\ln\frac{k_{23}k_{34}k_{45}k_{52}}{k_{32}k_{43}k_{54}k_{25}}=\varDelta\mu,
 \label{eqaff2multi4} 
\end{equation}
and
\begin{equation}
\A_3=\ln{k_{12}k_{24}k_{45}k_{51}}{k_{21}k_{42}k_{54}k_{15}}=0.
 \label{eqaff3multi4} 
\end{equation}
These three affinities determine the affinity of the two remaining three states 
cycles,which are
\begin{equation}
\A_4=\ln\frac{k_{23}k_{34}k_{42}}{k_{32}k_{43}k_{24}}=\varDelta\mu,
\end{equation}
and
\begin{equation}
\A_5=\ln\frac{k_{12}k_{25}k_{51}}{k_{21}k_{52}k_{15}}=0.
\end{equation}
The dominant cycle has affinity $\A_1=\varDelta\mu$ and five states. The numerical evidence for the bound 
$\R\le f(\Delta\mu,5)$ is shown in Fig. \ref{figmulti4b}.

For the multicyclic network given in Fig. \ref{fig2}  we have performed a similar analysis. 
This network has a total of 11 cycles. We identify the states $E,ES,EP,ESS,ESP,EPP$ in Fig. \ref{fig2}  
as $1,2,3,4,5,6$, respectively. We have generated two sets of points. The first set has 
$10^9$ points and we just accepted results fulfilling $\R\ge\tanh(\varDelta\mu/8)$. For this set we have 
chosen the rates $k_{12}$, $k_{31}$, $k_{45}$,  and $k_{56}$  as $\textrm{e}^{\varDelta\mu/6}10^{(2+x)/2}$;
the rates $k_{21}$, $k_{25}$, $k_{32}$, $k_{35}$, $k_{36}$, $k_{42}$, $k_{52}$, $k_{53}$, $k_{54}$, $k_{56}$, and
$k_{65}$ as $10^x$; with $x$ uniformly distributed between $-2$ and $2$. The four remaining rates $k_{13}$,
$k_{23}$, $k_{24}$, and $k_{63}$ were determined by the constraints set by the affinities. The second set has 
$10^7$ points and the $14$ independent rates were chosen as $10^x$, with $x$ uniformly distributed between $-2$ and $2$.

In summary,  we have confirmed our bound numerically for four different networks aside from the one displayed in Fig. \ref{fig2}. Since these 
networks do not share any symmetry, our full numerics provides strong evidence for our bound conjectured in Eq. \eqref{secondmain}. 

%==========================================================================
\section{Phosphorylation level in the model for a single KaiC}
%==========================================================================
\label{app4}

We show that the phosphorylation level of the KaiC protein displays oscillations, with 
the number of coherent oscillations characterized by the ratio $\R$. 

For the single KaiC model we analyze in Sec. \ref{sec4a}, the stochastic matrix reads
\begin{widetext}
\begin{equation}
\left(
\begin{array}{cccccccccccccc}
 -r_1 & e^{E/6} \gamma  & 0 & 0 & 0 & 0 & 0 & 0 & 0 & 0 & 0 & 0 & 0 & e^{E} k \\
 e^{\varDelta\mu /2} \gamma  & -r_2 & e^{E/6} \gamma  & 0 & 0 & 0 & 0 & 0 & 0 & 0 & 0 & 0 & e^{2 E/3} k & 0 \\
 0 & e^{\varDelta\mu /2} \gamma  & -r_2 & e^{E/6} \gamma  & 0 & 0 & 0 & 0 & 0 & 0 & 0 & e^{E/3} k & 0 & 0 \\
 0 & 0 & e^{\varDelta\mu /2} \gamma  & -r_2 & e^{E/6} \gamma  & 0 & 0 & 0 & 0 & 0 & k & 0 & 0 & 0 \\
 0 & 0 & 0 & e^{\varDelta\mu /2} \gamma  & -r_3 & e^{E/6} \gamma  & 0 & 0 & 0 & k & 0 & 0 & 0 & 0 \\
 0 & 0 & 0 & 0 & e^{\varDelta\mu /2} \gamma  & -r_4 & e^{E/6} \gamma  & 0 & k & 0 & 0 & 0 & 0 & 0 \\
 0 & 0 & 0 & 0 & 0 & e^{\varDelta\mu /2} \gamma  & -r_5 & k & 0 & 0 & 0 & 0 & 0 & 0 \\
 0 & 0 & 0 & 0 & 0 & 0 & e^{E} k & -r_1 & e^{E/6} \gamma  & 0 & 0 & 0 & 0 & 0 \\
 0 & 0 & 0 & 0 & 0 & e^{2 E/3} k & 0 & e^{\varDelta\mu /2} \gamma  & -r_2 & e^{E/6} \gamma  & 0 & 0 & 0 & 0 \\
 0 & 0 & 0 & 0 & e^{E/3} k & 0 & 0 & 0 & e^{\varDelta\mu /2} \gamma  & -r_2 & e^{E/6} \gamma  & 0 & 0 & 0 \\
 0 & 0 & 0 & k & 0 & 0 & 0 & 0 & 0 & e^{\varDelta\mu /2} \gamma  & -r_2 & e^{E/6} \gamma  & 0 & 0 \\
 0 & 0 & k & 0 & 0 & 0 & 0 & 0 & 0 & 0 & e^{\varDelta\mu /2} \gamma  & -r_3 & e^{E/6} \gamma  & 0 \\
 0 & k & 0 & 0 & 0 & 0 & 0 & 0 & 0 & 0 & 0 & e^{\varDelta\mu /2} \gamma  & -r_4 & e^{E/6} \gamma  \\
 k & 0 & 0 & 0 & 0 & 0 & 0 & 0 & 0 & 0 & 0 & 0 & e^{\varDelta\mu /2} \gamma  & -r_5 \\
\end{array}
\right)
\label{matdef}
\end{equation}
\end{widetext}
where $r_1\equiv k+\gamma\textrm{e}^{\varDelta\mu/2}$, $r_2\equiv k + \gamma\textrm{e}^{E/6}+\gamma\textrm{e}^{\varDelta\mu/2}$, $r_3\equiv
k\textrm{e}^{E/3}+\gamma\textrm{e}^{E/6}+\gamma\textrm{e}^{\varDelta\mu/2}$, $r_4\equiv k\textrm{e}^{2E/3}+\gamma\textrm{e}^{E/6}+\gamma\textrm{e}^{\varDelta\mu/2}$, and $r_5\equiv k\textrm{e}^{E}+\gamma\textrm{e}^{E/6}$. The first seven states 
are related to the inactive form of the protein, with state $1$ corresponding to $C_0$ and state $7$ corresponding to $C_6$. The last seven states are related to the active form of the protein, 
with state $8$ corresponding to $\tilde{C}_6$ and state $14$ corresponding to $\tilde{C}_0$.

\begin{figure}
\includegraphics[width=65mm]{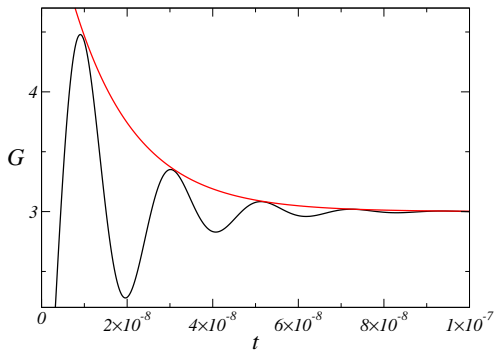}
\vspace{-2mm}
\caption{(Color online) Phosphorylation level of the KaiC protein. The parameters of the 
model are set to $k=\textrm{e}^{0.48097}$, $\gamma=\textrm{e}^{10.3475}$, $E=19.86$ and $\A=20$. 
Numerical calculation of the first eigenvalue gives $X_R\simeq 6.82\times10^7$ and $X_I\simeq 29.8\times10^7$,
which gives $\R\simeq 4.37$. The red solid line represents an exponential function with decay exponent $X_R$.
}
\label{figsup2} 
\end{figure}

The phosphorylation level of the protein is a state function. Its average is given by the expression 
\begin{equation}
G(t)=\sum_{i=0}^6 i[P_{C_i}(t)+P_{\tilde{C}_i}(t)],
\label{phosdef}
\end{equation}
$P_{C_i}(t)$ is the probability of configuration $C_i$ at time $t$. Initially the phosphorylation 
level is zero with the protein in state $C_0$, i.e.,  $P_{C_0}(0)=1$. The probabilities at time $t$
can be calculated with the expression
\begin{equation}
\mathbf{P}(t)=\exp(\mathbf{L}t)\mathbf{P}(0),
\label{evol}
\end{equation}
where $\mathbf{P}(0)= (1,0,0,0,0,0,0,0,0,0,0,0,0,0)^{T}$ and $\mathbf{L}$ given in Eq. \eqref{matdef}. We have calculated the 
phosphorylation level of the KaiC protein as a function of time from Eq. \eqref{phosdef}
and Eq. \eqref{evol}. The result is shown in Fig. \ref{figsup2}. Clearly the exponential decay of the amplitude 
and the period of oscillation are determined by the first eigenvalue of the matrix in Eq. \eqref{matdef} (see caption of Fig. \ref{figsup2}).

%==========================================================================
\section{Relation between $\R$ and entropy production}
%==========================================================================
\label{app5}

In this appendix, we discuss the relation between the ratio $\R$ and the entropy production from stochastic thermodynamics \cite{seif12}, which we denote by $\sigma$.

For a unicyclic network with uniform rates $k^+= e^{\A/N}k^-$ this entropy production is given by
\begin{equation}
\sigma= (\A/N)(\textrm{e}^{\A/N}-1)k_-.
\label{entun}
\end{equation}
For large $N$, we obtain $\R= \A/2\pi$, as in Eq. \eqref{firstmain}, and, from $X_I=(\textrm{e}^{\A/N}-1)k_-\sin(2\pi/N)$, we 
obtain $\varDelta Q\equiv 2\pi\sigma/X_I= \A$. If $\sigma$ is interpreted as the rate of heat dissipated to the environment \cite{seif12}, 
$\varDelta Q$ is the dissipated heat per period of oscillation. Hence, for a unicyclic network with large number of states $N$, we find 
\begin{equation}
\R^{-1}= 1/\varDelta Q.
\label{reltu}
\end{equation}
This expression is a particular case of the relation found in \cite{cao15}, which states that after some critical value $\Delta Q_c$, 
for which oscillations set in, the inverse of the number of biochemical oscillations decay to some plateau  
with $(\Delta Q-\Delta Q_c)^{-1}$. For our particular case both $\Delta Q_c$ and the plateau are zero.  
In \cite{cao15} this relation was demonstrated to be fulfilled for several different models.  

\begin{figure}
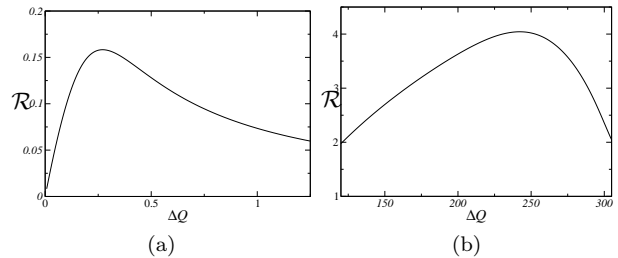

\subfigure[]{\psfrag{R}{$\mathcal{R}$}\includegraphics[width=40mm]{./fig11a.eps}\label{figsupact2a}}
\subfigure[]{\psfrag{R}{$\mathcal{R}$}\includegraphics[width=39mm]{./fig11b.eps}\label{figsupact2b}}
\vspace{-2mm}
\caption{Relation between energy dissipation and number coherent oscillations. (a) Unicyclic model with $N=3$, $k_1^-=k_2^-=k_3^-=1$, 
$k_1^+=k_3+= \textrm{e}^{\A/4}$ and $k_2^+= \textrm{e}^{\A/2}$. (b) Model for a single KaiC with $k=1$, $\gamma= \textrm{e}^5$, and $E=10$.}   
\label{figsupact2} 
\end{figure}

While this relation is true for an unicyclic network with uniform rates that maximize $\R$, for arbitrary rates $\R$ 
can also decrease with an increase in $\Delta Q$. The entropy production for the generic unicylic model in Eq. \eqref{fullreaction} reads
\begin{equation}
\sigma= \A(P_Nk_1^+-P_1k_1^-),
\end{equation}
where $P_i$ is the stationary probability of state $i$. As an example, we consider the unicyclic model for $N=3$ with $k_1^-=k_2^-=k_3^-=1$, 
$k_1^+=k_3^+= \textrm{e}^{\A/4}$ and $k_2^+= \textrm{e}^{\A/2}$. In Fig. \ref{figsupact2a} we show that $\R$ as a function of $\varDelta Q= 2\pi \sigma/X_I$ has a maximum,
where we vary the affinity $\A$. Therefore, the number of coherent oscillations can also decrease with an increase in energy dissipation. A similar behavior has been observed in \cite{qian00}, with the main 
difference that instead of varying $\varDelta\mu$ the authors vary the temperature. The maximum of $\R$ as a function of temperature was identified as stochastic resonance.

The same maximum was also observed for the single KaiC model, as shown in Fig. \ref{figsupact2b}. In this case, the entropy production can be written as
\begin{equation}
\sigma=\varDelta\mu\gamma\sum_{i=0}^{5}\left(\textrm{e}^{\varDelta\mu/2}P_{C_{i}}-\textrm{e}^{E/6}P_{C_{i+1}}\right),
\end{equation}
where $P_{C_{i}}$ is the stationary probability of state $C_i$. In this expression we used the Schnakenberg cycle decomposition of the entropy production \cite{seif12}.

%==========================================================================
\section{Activator-inhibitor model}
%==========================================================================
\label{app6}

In this appendix, we define the activator-inhibitor model from \cite{cao15}. 
The model has four different chemical species: the activator $R$, the inhibitor $X$, the 
enzyme $M$ and the phosphatase $K$. An enzyme $M$ can be in four different states, which form the
phosphorylation cycle
\begin{align}
& M+R+K+ATP\xrightleftharpoons[k_2^-]{k_1^+} MR+K+ATP\xrightleftharpoons[k_3^-]{k_2^+}\nonumber\\
& M_p+ADP+K+R\xrightleftharpoons[k_{4}^-]{k_{3}^+}M_pK+ADP+R\xrightleftharpoons[k_{1}^-]{k_{4}^+}\nonumber\\
& M+ADP+P_i+K+R.
\label{phoscyc} 
\end{align}
The concentrations of $ATP$, $ADP$ and $P_i$ are assumed to be fixed. From the generalized detailed balance relation, the rates in Eq. \eqref{phoscyc} fulfill
\begin{equation}
\textrm{e}^{\varDelta\mu}=k_1^+k_2^+k_3^+k_4^+/(k_1^-k_2^-k_3^-k_4^-) 
\end{equation}
where $\varDelta\mu$ is the free energy liberated in one $ATP$ hydrolysis. The activator $R$ catalyzes
the phosphorylation of the enzyme $M$ and the phosphatase $K$ catalyzes the dephosphorylation of $M$. 

The enzyme in the phosphorylated state $M_p$ catalyzes the creation of both the activator $R$ with rate $l_0$ and the inhibitor $X$ with rate $l_3$. 
The activator $R$ can also be spontaneously created with a rate $l_1$. The inhibitor $X$ catalyzes the degradation of $R$ with rate $l_2$ and can be 
spontaneously degraded with a rate $l_4$. Hence, we have the following chemical reactions
\begin{align}
& M_p\xrightleftharpoons[\epsilon]{l_0}M_p+R, \nonumber\\
& \emptyset\xrightleftharpoons[\epsilon]{l_1} R, \nonumber\\
& X+R\xrightleftharpoons[\epsilon]{l_2} X, \nonumber\\
& M_p\xrightleftharpoons[\epsilon]{l_3} M_p+X, \nonumber\\
& X\xrightleftharpoons[\epsilon]{l_4}\emptyset. 
\label{reacchem} 
\end{align}
These are equilibrium reactions and a cycle that involves only them must have zero affinity.
The only way to get a cycle with nonzero affinity in this model is to use the chemical reactions in Eq. \eqref{phoscyc}. 
The reversed rates $\epsilon$ are assumed to be very small so that we can set them to zero. Formally, they must be nonzero 
for thermodynamic consistency, however, in a numerical simulation we can just set them to zero, instead of using a very 
small $\epsilon$ that will lead to the same results.

The total number of enzymes $N_{M_T}=N_{M}+N_{M_pK}+N_{M_p}+N_{MR}$, and that of phosphatases $N_K$ are conserved. The number of activators $R$ fulfills $N_R\ge 1$ and the number of inhibitors $X$ 
fulfills $N_X\ge 1$. A state of the system is then determined by the vector $\mathbf{N}=(N_R,N_X,N_{M},N_{M_pK},N_{M_p},N_{MR})$. 
The volume of the system is written as $V$ and the concentration of the chemical species $X$, for example, is denoted by $n_X\equiv N_X/V$.   
The master equation that defines this model reads
\begin{widetext} 
\begin{align}
& \frac{d}{dt}P(\mathbf{N})= (l_0 N_{M_p}+l_1)P(N_R-1,\ldots)+  l_2 n_{X}(N_{R}+1)P(N_R+1,\ldots)\nonumber\\
& +(l_3 N_{M_p})P(\ldots,N_X-1,\ldots)+  l_4(N_{X}+1)P(\ldots,N_X+1,\ldots)\nonumber\\
& +k_1^-(N_M+1)(n_K-n_{M_pK}+\delta)P(\ldots,N_M+1,N_{M_pK}-1,\ldots)\nonumber\\
& +k_1^+(N_M+1)(n_R-n_{MR}+\delta)P(\ldots,N_M+1,\ldots,N_{M_R}-1)\nonumber\\
& +k_2^-(N_{MR}+1)P(\ldots,N_{M}-1,\ldots,N_{MR}+1)+k_2^+(N_{MR}+1)P(\ldots,N_{M_p}-1,N_{MR}+1)\nonumber\\
& +k_3^-(N_{M_p+1})(n_R-n_{MR}+\delta)P(\ldots,N_{M_p}+1,N_{MR}-1)+k_3^+(N_{M_p}+1)(n_K-n_{M_pK}+\delta)P(\ldots,N_{M_pK}-1,N_{M_p}+1,\ldots)\nonumber\\
& +k_4^-(N_{M_pK}+1)P(\ldots,N_{M_pK}+1,N_{M_p}-1\ldots)+k_4^+(N_{M_pK}+1)P(\ldots,N_M-1,N_{M_pK}+1,\ldots),
\end{align}
\end{widetext}
where $P(\mathbf{N})$ is the probability to be in state $\mathbf{N}$ at time $t$ and $\delta\equiv1/V$. Note that the rates 
$l_2$, $k_1^{\pm}$, $k_3^{\pm}$ have dimension $V^{-1}t^{-1}$, whereas the other rates have dimension $t^{-1}$. 
In the above equation, we set to zero the 
probability of configurations that violate the constraints $N_R\ge 1$, $N_X\ge 1$, $N_{M_T}=N_{M}+N_{M_pK}+N_{M_p}+N_{MR}$ and $N_K\ge N_{M_pK}$.

We have performed continuous time Monte Carlo simulations of this model and calculated the correlation function
\begin{equation}
C(t)\equiv \langle (N_X(t)-\langle N_X\rangle)(N_X(0)-\langle N_X\rangle) \rangle,
\end{equation}
where the brackets denote an average over stochastic trajectories and $\langle N_X\rangle$ is the average number of $X$ in the stationary state. The initial condition 
$N_X(0)$ was sampled from the stationary distribution: in a simulation we let the system reach the stationary state before the time $t=0$.

\begin{figure}
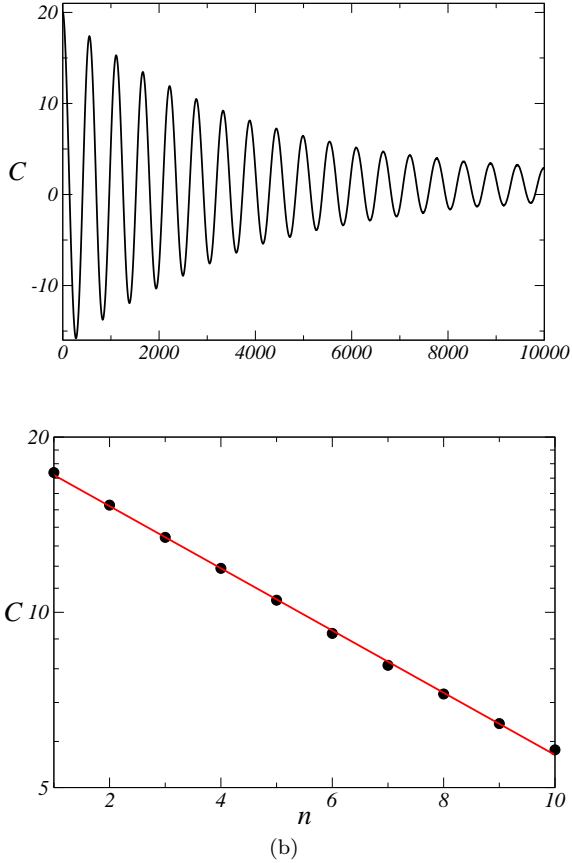

\subfigure[]{\includegraphics[width=75mm]{./fig12a.eps}\label{figsupacta}}
\subfigure[]{\includegraphics[width=75mm]{./fig12b.eps}\label{figsupactb}}
\vspace{-2mm}
\caption{(Color online) Numerical simulation of the activator-inhibitor model. Parameters are set to
$\varDelta\mu=12$ and $N_K=30$. The variable $n$ in (b) labels the the peak 
in (a). The solid red line in (b) is an exponential fit to the data that gives $\R\simeq 2\pi/0.12301\simeq 51.1$.}   
\label{figsupact} 
\end{figure}

The oscillating correlation function is shown in Fig. \ref{figsupacta}. 
The results presented in Fig. \ref{fig4} were obtained in the following way. We have adjusted an exponential function to the peaks of the oscillation as shown in Fig. \ref{figsupactb}. 
The exponent gives $T/\tau$, where $T$ is the period of oscillation and $\tau$ the decay time. The ratio $\R$ was then estimated as $\R= 2\pi \tau/T$.
The parameters of the model are similar to the parameters used in \cite{cao15}. They were set to $V=50$, $N_{M_T}=10V$, $l_0=l_2=l_3= 1$, $l_1=0.4$, $l_4=0.5$, $k_2^+=k_4^+=k_2^-=k_4^-=15$, $k_1^+=k_3^+=100$, and
$k_1^-=k_3^-=100\textrm{e}^{-\varDelta\mu/2}$. For the number of phosphatases we took $N_K= 30, 40, 50$ and for the driving force $\varDelta\mu=10, 12, 14, 16, 18, 20$.

\begin{figure}
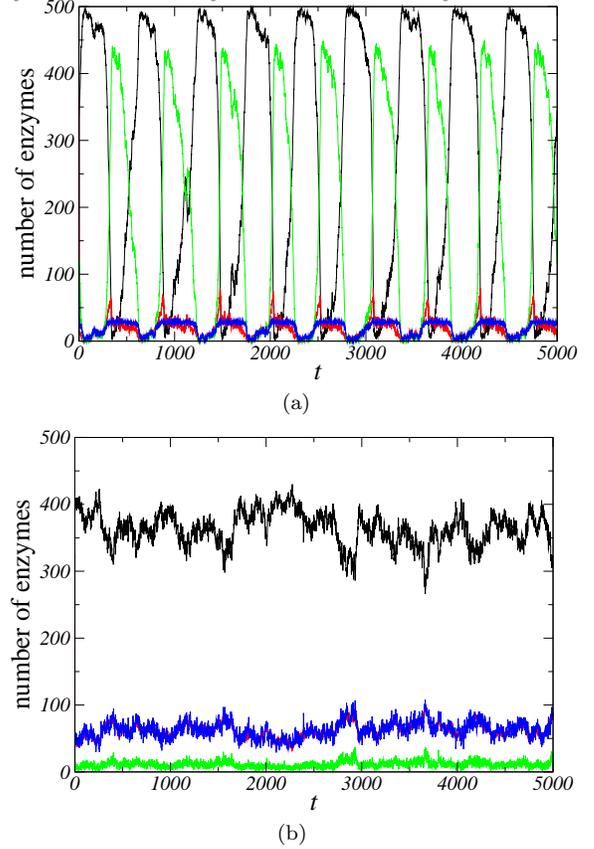

\subfigure[]{\includegraphics[width=75mm]{./fig13a.eps}}
\subfigure[]{\includegraphics[width=75mm]{./fig13b.eps}}
\vspace{-2mm}
\caption{(Color online) Stochastic trajectories of the activator-inhibitor model. The black trajectory shows
the number of enzymes in state $M$, the red in state $MR$, the green in state $M_p$, and 
the blue in state $M_pK$. Parameters are set to $\varDelta\mu=20$ and for (a) $N_K=30$ and (b) $N_K=150$.}   
\label{figtimes} 
\end{figure}

An important aspect of this model from Sec. \ref{sec4b} is that the competition for a small number of phophatase $K$ 
synchronizes the cycles of different enzymes $M$. If $N_K$ is too small we have no oscillations in $M_p$. If $N_K$ is
too large, the lack of competition for the phosphatase $K$ hinders oscillations in $M_p$. This feature is demonstrated 
in Fig. \ref{figtimes}, where we show two time series of the four different states of the enzyme for $N_K=30$ and $N_K=150$.
For $N_K=30$ we see clear oscillations, with the number $N_{M_pK}$ oscillating roughly between $0$ and $30$. Whenever  $N_{M_pK}(t)=30$,
there is no phosphatase left and several enzymes get stuck in state $M_p$, synchronizing the phosphorylation cycles of these enzymes. 
For $N_K=150$, the number $N_{M_pK}$ stays below $150$. Hence, there is always free phosphatase in the system, resulting in no synchrony
between different enzymes.

%==========================================================================
% References
%==========================================================================

%\bibliographystyle{aipnum4-1}
%\bibliographystyle{revtex}
%\bibliography{refs} 
%\providecommand*{\bibinfo}[2]{#2}
%\providecommand*{\eprint}[1]{#1}
%\providecommand*{\url}[1]{#1}

%

\end{document}